\documentclass[aps,prl,reprint,twocolumn,superscriptaddress]{revtex4-1}
\usepackage{graphicx,hyperref,pdfpages}
\pdfoutput=1
\begin{document}
%\title{Universal minimum superfluid response \& quantum phase transitions in two-dimensional superconductors }
%\title{Measurements of the low-frequency ac conductivity of disordered two-dimensional $\mathrm{Mo_{43}Ge_{57}}$ and $\mathrm{InO_x}$ superconducting films: evidence for a universal minimum superfluid response}
\title{Evidence for a universal minimum superfluid response in field-tuned disordered superconducting films measured using low frequency ac conductivity}

\author{S. Misra}
\affiliation{Joseph Henry Laboratories and Department of Physics, Princeton University}
\author{L. Urban}
\affiliation{Joseph Henry Laboratories and Department of Physics, Princeton University}
\affiliation{Department of Physics, University of Illinois at Urbana-Champaign}

\author{M. Kim}
\affiliation{Department of Physics, University at Buffalo, The State University of New York}

\author{G. Sambandamurthy}
\affiliation{Department of Physics, University at Buffalo, The State University of New York}

\author{A. Yazdani}
\email{yazdani@princeton.edu}
\affiliation{Joseph Henry Laboratories and Department of Physics, Princeton University}

\date{\today}

\begin{abstract}
Our measurements of the low frequency ac conductivity in strongly disordered two-dimensional films near the magnetic field-tuned superconductor-to-insulator transition show a sudden drop in the phase stiffness of superconducting order with either increased temperature or magnetic field. Surprisingly, for two different material systems, the abrupt drop in the superfluid density in a magnetic field has the same universal value as that expected for a Berezinskii-Kosterlitz-Thouless transition in zero magnetic field. The characteristic temperature at which phase stiffness is suddenly lost can be tuned to zero at a critical magnetic field, following a power-law behavior with a critical exponent consistent with that obtained in previous dc transport studies on the dissipative side of the transition.
\end{abstract}

\maketitle
In two dimensional (2D) systems, tuning localization by increasing disorder or applying a magnetic field drives a zero-temperature quantum phase transition between superconducting and insulating ground states \cite{GoldmanReview}. Despite decades of effort, uncovering a variety of novel phenomena, including the discovery of the scaling of transport properties \cite{HebardPRL1990,GoldmanPRL1991,YazdaniPRL1995,SteinerPRB2008}, unusual intervening zero temperature metallic phases \cite{SteinerPRB2008,VallesPRB2000,MasonPRL1999,YoonPRL2006}, and insulators with localized pairs \cite{SambandamurthyPRL2004, BehniaNature2006, VallesScience2007,Sacepe2011}, the underlying mechanism for the superconductor-insulator transition continues to be debated \cite{FisherPRL1990,DubiNature2007}.  Experimentally, efforts to probe this transition have focused almost exclusively on DC transport measurements that probe the samples once they are already strongly dissipative.  In contrast, ac conductivity measurements can probe the superconducting response of the system and directly detect the loss of superfluid-like response near the quantum phase transition out of the superconducting state, hence providing important complementary information. 

In the absence of a magnetic field, ac measurements have played a key role in demonstrating that the loss of superconducting response in 2D superconductors with increasing temperature occurs below the mean-field transition temperature, $T_{C0}$, at the Berezinskii-Kosterlitz-Thouless (BKT) transition temperature, $T_{BKT}$ \cite{BerezinskiiJETP,KT1973}. This transition occurs due to the unbinding of thermally generated vortex-anti-vortex pairs and is accompanied by a universal drop in the superfluid response of the system at $T_{BKT}$, which can be detected by ac measurements of the kinetic inductance \cite{HebardPRL1980, LembergerPRB2001}. In relatively clean thin films, the application of a magnetic field gives rise to a pinned Abrikosov lattice, the melting of which through a dislocation-unbinding transition can also be detected using ac inductive measurements sensing sudden loss of lattice rigidity \cite{YazdaniPRL1993}.  To date, the application of ac techniques to studying the superconductor-insulator transition in disordered systems has been limited to very high frequencies (very short length scales), where they have proven useful in detecting the remnant of superconducting correlations in the insulating phase \cite{ArmitagePRB2007}. 

In this letter, we use measurements of the low frequency ac conductivity to probe the loss of superconducting response in strongly disordered 2D films of $\mathrm{Mo_{43}Ge_{57}}$ and $\mathrm{InO_x}$ as they are tuned close to a quantum phase transition out of the superconducting state with the application of a magnetic field. Our main experimental finding is that the loss of superconductivity as a function of magnetic field at a fixed temperature occurs via a universal drop in the superfluid response, with a value corresponding to that predicted for the BKT transition in zero field. Furthermore, we show that the temperature at which the superfluid response is suddenly lost can be tuned to zero at a critical value of the field, following a power-law behavior with a critical exponent consistent with that obtained in previous dc transport studies of the dissipative side of the transition \cite{HebardPRL1990, GoldmanPRL1991, YazdaniPRL1995, SteinerPRB2008}. Our results suggest that driving the energy scale associated with the minimum superfluid response to zero with a magnetic field results in the quantum transition out of the superconducting state, and into either an unusual zero temperature metallic phase or an insulator.

Typical measurements of the complex conductance and kinetic inductance using our two-coil mutual inductance technique  \cite{JeanneretAPL1989} are shown Figure 1 for $\mathrm{Mo_{43}Ge_{57}}$ samples in zero magnetic field. Figure 1a shows the in-phase and out-phase response of our pick-up loop in response to shielding currents excited in the thin film sample by a drive coil, along with DC resistivity measurements on the same sample. The pickup voltages can be numerically transformed to obtain the real and imaginary parts of the sheet conductance $G(\omega)$ \cite{JeanneretAPL1989}, from which we compute the sample sheet impedance $Z(\omega) = R(\omega) + i\omega L(\omega)$, where $R(\omega)$ is the ac resistance and $L(\omega)$ is the ac inductance of the sample. The inverse inductance $L^{-1} = n_se^2/m$ , shown for example in Figure 1b for two different MoGe films in zero field, is proportional to the superfluid density $n_s$ ($m$ is the Cooper pair mass) and constitutes a direct measure of the phase stiffness of the superconducting order parameter in our samples.

Measurements in zero magnetic field demonstrate that our 2D samples undergo the expected vortex-anti-vortex unbinding BKT transition at which there is an abrupt loss of superfluid response at $T_{BKT} < T_{C0}$. As shown in Figure 1b, increasing the temperature results in a continuous decline in $L^{-1} = -\|\omega G\|^2/\Im[\omega G]$, which is well-defined up to the point where $\Im[\omega G]$ vanishes below our experimental noise floor ($\sim$.0005 $\mathrm{nH^{-1}}$). At this temperature, we find $L^{-1} \sim  n_s$ to approach a finite value just before superconductivity is lost in our sample. This behavior is similar to previous measurements of the superfluid density in other superconducting thin films \cite{HebardPRL1980, LembergerPRB2001}, as well as in two-dimensional superfluid films \cite{BishopPRL1978}, and trapped Bose gases \cite{DalibardNature2006}. The sudden drop in the superfluid density is predicted to have a universal value, independent of microscopic details \cite{KosterlitzPRL1977}, which for superconducting thin films can be expressed as $L^{-1}(T_{BKT}) = n_s(T_{BKT})e^2/m = (8\pi/\Phi_0^2)k_B T_{BKT}$ (corresponding to a line with a slope of 0.081 nH$\mathrm{^{-1}}$/K as shown in Figure 1b). The sudden change in our measured $L^{-1}$, shown in Figure 1b for two different MoGe films at low frequency, corresponds well to this predicted BKT universal jump and confirms that vortex anti-vortex unbinding underlies the loss of superconductivity in our samples in zero field with increasing temperature. Measurements at higher frequencies further confirm the shifting of the temperature at which $L^{-1}$ vanishes toward the mean field transition $T_{C0}$, as expected for a BKT transition \cite{Supplement}. 

\begin{figure}
\includegraphics[width=235pt]{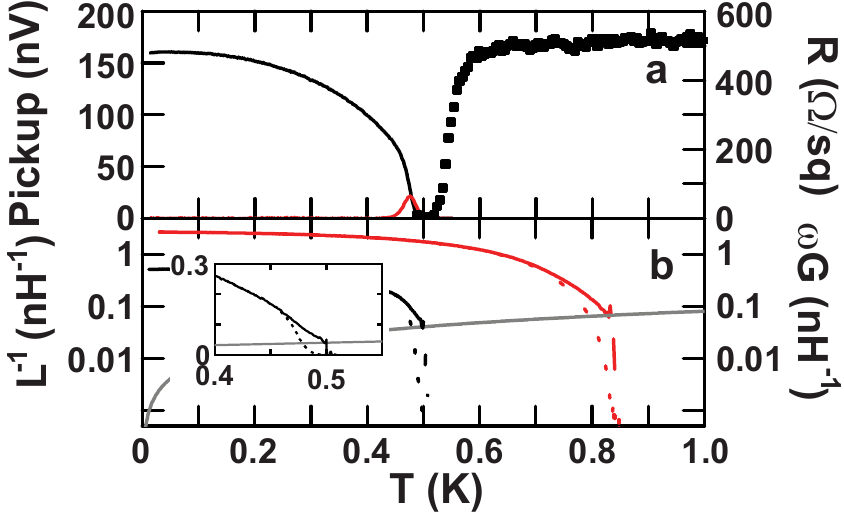}%
\caption{(a) The in-phase (black) and out-of-phase (red) voltage on the pickup coil ($f$ = 20 kHz and $I$ = 20 $\mathrm{\mu A}$ on the drive coil), plotted along with the resistivity from conventional dc transport (black squares), is shown as a function of temperature at zero field for a MoGe film. (b) Plotted here is $L^{-1}$ derived from the data in MoGe film in part (a) (black line), and a second less disordered film (red). Also shown are the imaginary part of $\omega G$ (dashed lines) and the BKT prediction for $L^{-1}$ (gray). The inset shows a close-up of the data from the more disordered film on a linear scale.}
\label{Figure 1}
\end{figure}

\begin{figure}
\includegraphics[width=235pt]{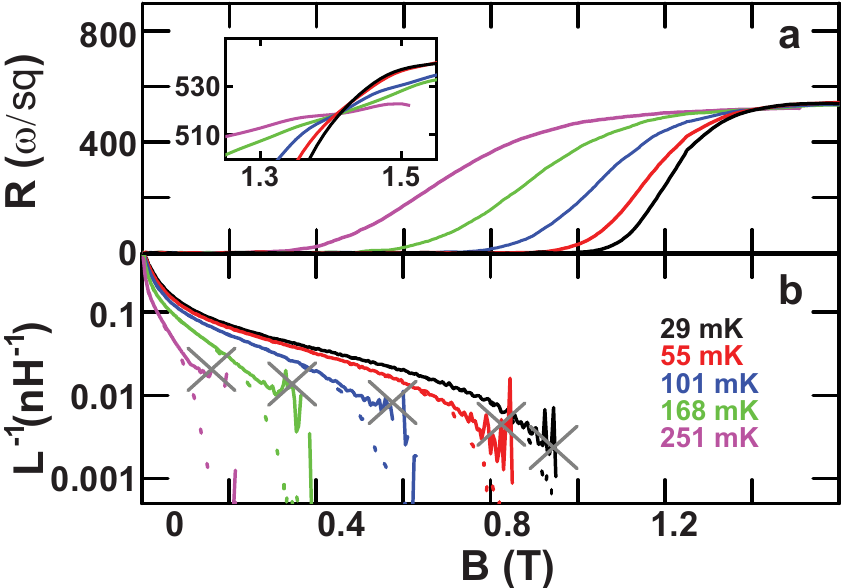}%
\caption{(a) Measurements of dc resistivity isotherms, shown here for the MoGe sample in Figure 1(a) and taken at the indicated temperature, shown as a function of field. (inset) A close-up indicates that the isotherms cross at a field of $B_X = \mathrm{1.41 \pm 0.02 T}$. (b) Measurements of $L^{-1}$ isotherms for the same film as (a) taken at 50 kHz show a discontinuous jump which moves to larger values of the field at lower temperatures. Also shown are the imaginary part of $\omega G$ (dashed lines) and the BKT prediction for $L^{-1}$ (gray crosses). }
\label{Figure 2}
\end{figure}

The application of a magnetic field strongly suppresses superconductivity in thin films, eventually driving the system through a quantum phase transition out of the superconducting state. Typically, dc resistivity measurements have been used to identify a critical value of the magnetic field $B_X$, where resistance isotherms cross each other and around which such data scales in a manner consistent with theoretical work on the superconductor-insulator transition \cite{HebardPRL1990,GoldmanPRL1991,YazdaniPRL1995,SteinerPRB2008}. However, resistivity measurements have also found a flattening of the resistance at the lowest temperatures in a field range close to $B_X$ \cite{VallesPRB2000,MasonPRL1999}, introducing the possibility of a metallic phase over an intervening range of magnetic field as the samples are driven from superconducting to insulating ground states. It is still debated whether such behavior is due to the lack of cooling of the samples rather than true metallic behavior. While such transport measurements, as shown for example in Figure 2a for our MoGe sample, directly probe the changes of dissipation with magnetic field, they do not probe the loss of superfluid response. To obtain such information, we turn to our measurements of $L^{-1}$ from the ac mutual inductance technique performed on the same sample, as shown in Figure 2b. 

In general, application of a magnetic field alters the inductive response of a superconductor not only through the suppression of the superfluid density, but also through field-induced vortices. While detailed modeling can be used to account for both these effects in $L^{-1}$ \cite{Supplement},  we focus instead on the point where the superfluid-like response is lost in our samples with increasing magnetic field. Remarkably, we find that at each temperature there is a specific value of field at which $L^{-1}$ shows a precipitous drop to zero. The sudden loss of $L^{-1}$ with field coincides with a sudden increase in dissipation that is first detected in the ac resistivity, and eventually can also be measured in dc transport \cite{Supplement}. The most intriguing aspect of the sudden loss of the superfluid response is that the value of the sudden drop corresponds to the same value for the universal jump in the superfluid density for the zero field BKT transition (crosses in Figure 2b). Moreover, with decreasing temperature, the characteristic field for the sudden change in superfluid-like response is continuously shifted to higher field.

\begin{figure}
\includegraphics[width=235pt]{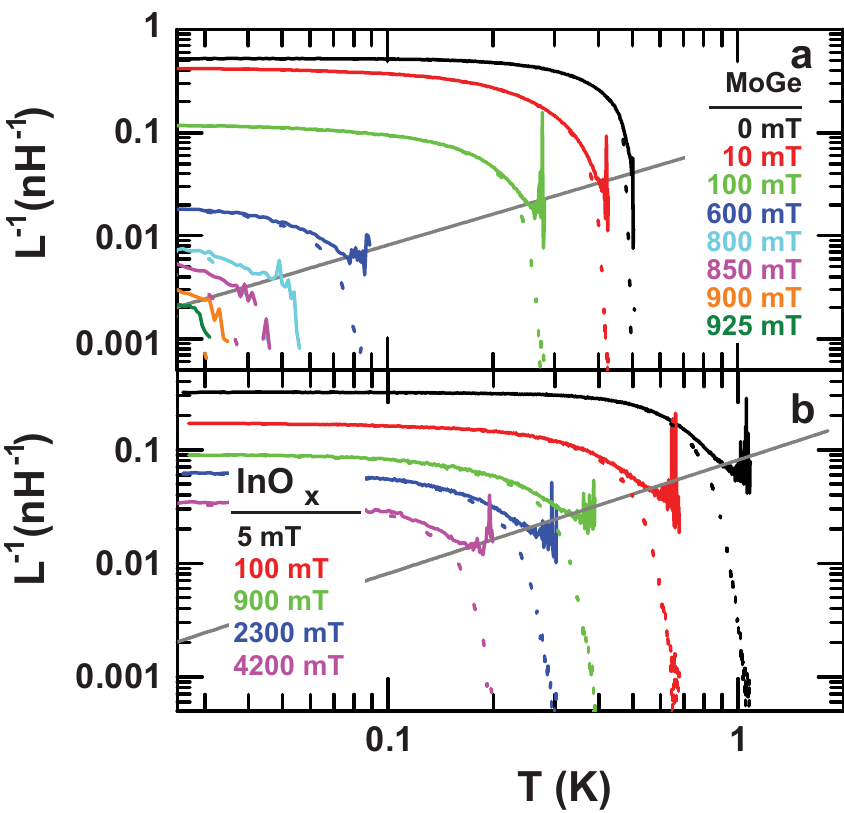}%
\caption{Temperature sweeps of $L^{-1}$, taken here at 20 kHz, in the presence of an applied magnetic field show a discontinuous jump to zero in both (a) MoGe and (b) $\mathrm{InO_x}$ thin film samples. Also shown are the imaginary part of $\omega G$ (dashed lines) and the BKT prediction for $L^{-1}$ (gray).}
\label{Figure 3}
\end{figure}

A more accurate determination of the evolution of the sudden loss of superfluid response can be obtained by measuring the temperature dependence of $L^{-1}$ at fixed values of the magnetic field. As shown in Figures 3a and 3b, such measurements on both MoGe and $\mathrm{InO_x}$ thin films show that the application of a field reduces the overall superfluid response of the sample at all temperatures, while increasing temperature results in a sudden loss of response abruptly at a temperature $T^*$.  Our main experimental finding is that the jump in the superfluid density of strongly disordered 2D samples, whether field-tuned at a given temperature (Figure 2b) or temperature-tuned in the presence of a magnetic field (Figures 3a and 3b), follows the universal BKT value, independent of material system. Thus, surprisingly, the minimum strength of superfluid stiffness is determined by the zero-field BKT criterion, even in the presence of a magnetic field that introduces a sizable population of vortices. While the size of the jump becomes increasingly more difficult to measure at higher fields as $T^*$ gets smaller, the size of the jump continues to follow the BKT criterion approaching the quantum phase transition.  

It is important to recognize that the loss of superfluid response in our samples is distinct from earlier studies in cleaner 2D MoGe samples, which showed melting of the Abrikosov lattice via a dislocation-anti-dislocation unbinding transition \cite{YazdaniPRL1993}. Notably, the details associated with the pinning of vortices would be expected to differ between the two samples studied here, while a universal behavior of the superfluid response is measured. In addition, the loss of superfluid response at the 2D vortex lattice melting transition in cleaner thin films, although sudden, does not occur with the universal BKT value \cite{YazdaniPRL1993}. Finally, as shown previously, increasing the disorder through decreasing film thickness, and hence suppressing the zero-field superconducting transition temperature significantly compared to thick (3D) films, results in the eventual disappearance of the experimental signatures of vortex lattice melting since pinning and creep dominates the behavior of a glassy vortex system \cite{YazdaniPRB1994}. Not only do the MoGe and $\mathrm{InO_x}$ samples examined here have zero-field superconducting transitions which are dramatically suppressed in comparison to thicker films (for example, 500 mK for the MoGe film here, compared to 1.05 K for a thicker film close to bulk $T_c$), but they also show a frequency dependence at finite field which is consistent with vortex creep \cite{Supplement}, both of which indicate they should have an extremely disordered vortex lattice. The superfluid response of such disordered thin films has never been previously probed at low frequencies approaching the field-tuned quantum phase transition out of the superconducting state. 

\begin{figure}
\includegraphics[width=235pt]{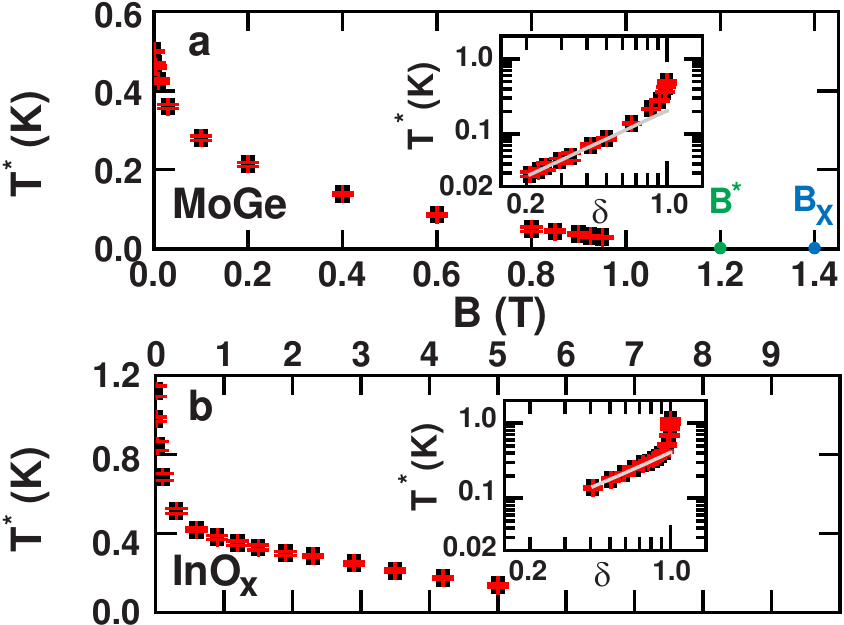}%
\caption{Shown are the temperatures ($T^*$), along with error bars (red lines), at which $L^{-1}$ is measured to jump to
zero in our temperature sweeps at fixed field in the (a) MoGe and (b) $\mathrm{InO_x}$ samples (f = 20
kHz). The inset shows a fit of the data to the power law form $T^* \sim \delta ^{\nu z}$, where $\delta = (1 - B/B^*)$ on a log-log scale. In part (a), the fitted critical field $B^*$ from the ac measurements is shown in green, along with the crossing field $B_X$ from the resistivity data of Figure 2 in blue.}
\label{Figure 4}
\end{figure}

To determine the connection between the loss of superfluid response and possible quantum phase transitions in our strongly disordered thin films, we examine the behavior of $T^*$ as a function of magnetic field. As shown in Figure 4, the temperature $T^*$ at which superfluid response is lost extrapolates to zero, suggesting the presence of a quantum phase transition at a critical value of the magnetic field where superconducting behavior is lost at zero temperature. We find that simple extrapolation of the data, shown in Figure 4, to zero temperature finds a critical field ($B^* = \mathrm{1.2 \pm 0.1 T}$) that, despite a significant error bar, is smaller than the crossing field extracted from resistance isotherms such as those shown in the inset of Figure 2 ($B_X = \mathrm{1.41 \pm 0.02 T}$). Limited frequency-dependent data taken at our base temperature imply that the superconducting phase could terminate at fields even lower than $B^*$ in the zero-frequency limit \cite{Supplement}. This discrepancy between $B^*$ and $B_X$ may in fact be due to the presence of an intervening metallic phase that has no superfluid response, but has finite resistance even in the limit of zero temperature \cite{MasonPRL1999}. We proceed to explore if the vanishing $T^*$ has properties consistent with that of a quantum critical point, near which we anticipate that $T^* \sim (1 - B/B^*) ^{\nu z}$, where $\nu z$ are  exponents governed by the critical fluctuations near the quantum phase transition \cite{SondhiRMP1997,VojtaRPP2003}. The inset of Figure 4 shows the extracted value of $T^*$ as a function of magnetic field, and the corresponding power-law fits near where $T^*$ approaches zero to obtain $B^*$ ($\mathrm{1.2 \pm 0.1 T}$ for MoGe and $\mathrm{8.5 \pm 1.5 T}$ for $\mathrm{InO_x}$) and the combination of critical exponents $\nu z$ ($\mathrm{1.25 \pm 0.25}$ for MoGe and $\mathrm{1.3 \pm 0.4}$ for $\mathrm{InO_x}$). It is interesting to note that, despite significant error bars, these critical exponents are consistent with transport studies on similar MoGe and $\mathrm{InO_x}$ samples, when these studies have limited their analysis to exclude resistivity data showing finite dissipation extrapolating to zero temperature \cite{HebardPRL1990, YazdaniPRL1995, MasonPRL1999, SteinerPRB2008}.

Our measurements demonstrate that the loss of superfluid response in a disordered 2D superconductor in the presence of a field is surprisingly well-described by the BKT criterion for minimum sustainable superfluid response familiar from other 2D superfluid-insulator or superfluid-normal transitions, despite the presence of a net vorticity resulting from the external magnetic field. Following the characteristic jump in the superfluid response with field and temperature, we arrive at the conclusion that the energy scale associated with this minimum superfluid response is driven to zero at a critical value of the magnetic field. In the absence of a magnetic field, the idea that a vortex-anti-vortex unbinding mechanism, such as that demonstrated for our films at finite temperature (Figure 1), can underlie the quantum phase transition out of the superconducting state has been previously considered when the transition is driven by disorder  \cite{FisherPRL1990}. Increased disorder suppresses the overall superfluid density, while the minimum sustainable superfluid density is still described by the vortex-anti-vortex unbinding criterion, resulting in a continuous tuning of $T_{BKT}$ to zero. Extending this interpretation to our results is complicated by the simple fact that the applied field changes the energetics of vortices versus anti-vortices, and one usually considers the loss of superconductivity in the context of melting of the Abrikosov vortex lattice via dislocation-anti-dislocation unbinding rather than vortex-anti-vortex unbinding. In contrast, our experiments are in the limit of a strongly disordered vortex lattice, where pinning and creep of vortices in a glassy state dominates over melting phenomena associated with that of a clean system.

A key question our experiments raise is whether mechanisms other than vortex-anti-vortex unbinding can result in a minimum superfluid density criterion similar to that of the BKT transition. Understanding the universality of our minimum sustainable superfluid response in the strongly disordered samples studied here could provide the context to have a unified explanation of the destruction of superconductivity at zero temperature both as a function of disorder and magnetic field. 

\bibliography{paper}
\clearpage


\section*{Supplementary material}

# Supplementary Information: Evidence for a universal minimum superfluid response in field-tuned disordered superconducting films measured using low frequency ac conductivity

S. Misra,[1] L. Urban,[1, 2] M. Kim,[3] G. Sambandamurthy,[3] and A. Yazdani[1, *]

[1]*Joseph Henry Laboratories and Department of Physics, Princeton University*
[2]*Department of Physics, University of Illinois at Urbana-Champaign*
[3]*Department of Physics, University at Buffalo, The State University of New York*
(Dated: December 5, 2012)

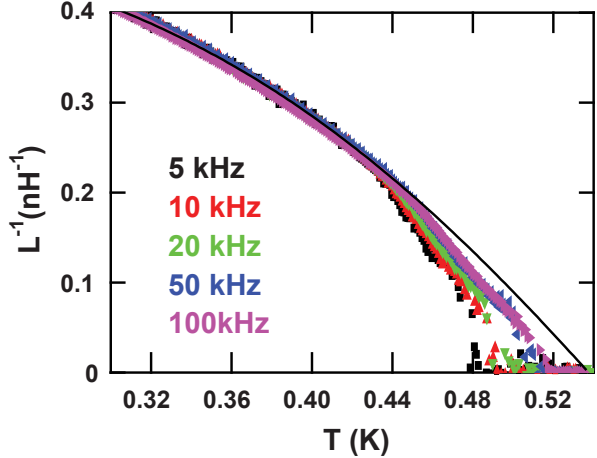

FIG. S1. Frequency dependence in zero field. The inverse sheet inductance $L^{-1}$ of the MoGe film as a function of temperature in zero magnetic field for five frequencies is shown in colored symbols. The solid black line is the mean-field BCS prediction for a superconductor in the dirty limit.

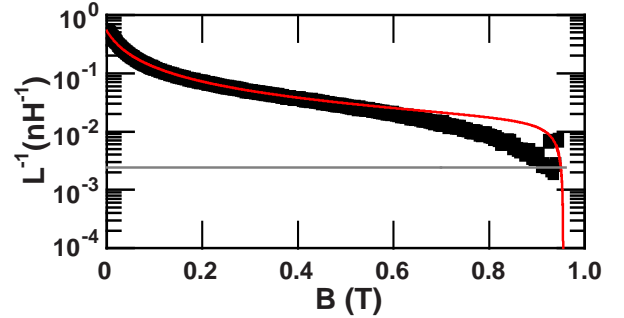

FIG. S2. Field dependence at low temperature. The inverse sheet inductance $L^{-1}$ of the MoGe film measured at 50 kHz is shown as a function of magnetic field at a temperature of 30 mK as black squares. The grey line is the BKT prediction for the jump in $L^{-1}$ at this temperature. The red line is a fit to the Coffey- Clem model of vortex motion [4] with parameters $\lambda(0) = 3.86 \mu m$, $\alpha = 3.6 N/m^2$, and $\eta = 0.25 \mu N - s/m^2$. Note that the model has more free parameters than is necessary to fit just the inverse inductance, and that a reasonable value of $\alpha$ and $\eta$ was taken by trying to match the magnitude of the ac resistivity (data shown in Figure S3).

## I. FREQUENCY DEPENDENCE OF $L^{-1}$ NEAR SUPERCONDUCTING TRANSITION IN ZERO FIELD

Measurements of the ac sheet impedance of a superconductor are necessarily frequency dependent in zero applied magnetic field due to the thermal generation and unbinding of vortex-anti-vortex pairs. In general, we expect that the frequency of our ac measurements will probe the sample at different length scales, with lower frequencies corresponding to longer length scales. In zero magnetic field, the temperature dependence of the amplitude of the order parameter should result in a suppression of the inverse inductance as $\frac{L^{-1}(T)}{L^{-1}(0)} = \sqrt{cos\left(\frac{\pi t^2}{2}\right)} tanh\left[\frac{1.76}{2t}\sqrt{cos\left(\frac{\pi t^2}{2}\right)}\right]$, where we have started from the BCS expression for the penetration depth $\lambda$ in the dirty limit, taken $L^{-1} \sim \lambda^{-2}$, and assumed that the amplitude of the order parameter is given by $\Delta(T) = 1.76 k_B T_C \sqrt{cos\left(\frac{\pi t^2}{2}\right)}$, in which $t = T/T_C$. This should occur uniformly throughout the sample, and hence be frequency independent. As seen in Figure S1, below 0.43 K, $L^{-1}$ is independent of frequency, and is

fit reasonably well by the BCS prediction between 0.31 K and 0.43 K. Above this temperature, the BCS prediction breaks down, as it does not account for the aforementioned unbinding of vortex-anti-vortex pairs. On increasing the temperature, vortex-anti-vortex pairs at the largest separations will be the first to unbind [1], resulting in the loss of phase coherence and a complete suppression of $L^{-1}$ first at lower frequencies. More thermal energy is required to unbind those at shorter separations, meaning $L^{-1}$ will go to zero at higher temperatures for data taken at higher frequencies, which probe a shorter length scale [2, 3].

## II. THE INFLUENCE OF FIELD-INDUCED VORTICES ON $L^{-1}$ AT LOW TEMPERATURES.

The inverse inductance in the presence of a magnetic field decreases not only from the suppression of the kinetic response of the superconducting pairs, but also from the presence and motion of vortices. We find that the low temperature ($T < T^*$) behaviour of $L^{-1}$ in an

applied field is captured well by a model for the sheet impedance proposed by Coffey and Clem [4]. They calculate a generalized complex ac penetration depth $\lambda_{ac}^2 = \lambda^2 + \frac{1}{\lambda_C^{-2} + 2i\delta_F^{-2}}$, which includes the kinetic response of the superfluid through the bulk penetration depth $\lambda^2 = \frac{\lambda^2(0)}{(1-B/B_C)}$, and incorporates the effect of the restoring force from vortex pinning ($\alpha_L$) through the Campbell penetration depth $\lambda_C^2 = \frac{B\Phi_0}{\mu_0\alpha_L}$, and the vortex viscous drag ($\eta$) through the flux-flow diffusion length $\delta_F^2 = \frac{B\Phi_0}{\mu_0\omega\eta}$ (here, $B$ is the applied magnetic field, $B_C$ is the field where $L^{-1}$ vanishes, $\Phi_0$ is the flux quantum, $\mu_0$ is the magnetic permeability of vacuum, and $\omega$ is the frequency). The complex sheet impedance for a 2D film is then given by $Z(\omega) = \frac{i\omega\mu_0\lambda_{ac}^2}{d}$, where $d$ is the thickness of the film. As can be seen in Figure S2, the model fits the measured $L^{-1}$ data well, with the suppression of the bulk penetration depth being largely responsible for the behavior near zero field, and the motion of pinned vortices being largely responsible for the behavior in finite field up to $\sim 0.6$ T.

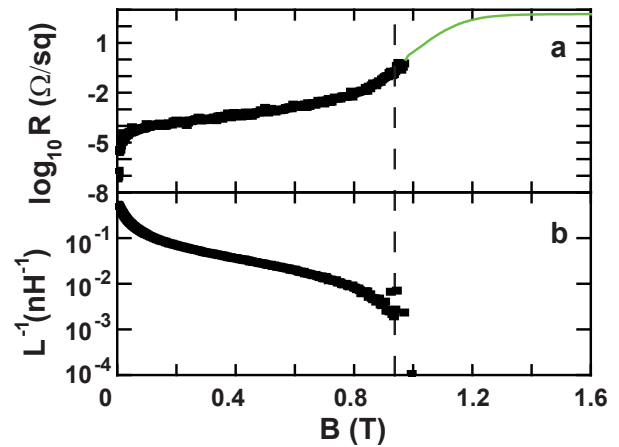

FIG. S3. The ac and dc resistivity near the jump in $L^{-1}$. (a) The resistivity of the MoGe sample measured at 30 mK measured using the coil at a frequency of 50 kHz (solid squares) and conventional dc transport (green line). (b) The inverse sheet inductance acquired simultaneously with the ac resistivity shown in (a). The vertical dashed line is the field at which $L^{-1}$ drops discontinuously to zero.

## III. BEHAVIOUR OF RESISTIVITY AT LOW TEMPERATURE

Our ac coil measures a small, but finite, amount of dissipation in finite field even at the lowest temperatures (30 mK) due to the presence and motion of vortices, as discussed in Supplementary Section II. As can be seen in Figure S3, the sudden drop in $L^{-1}$ at 0.94 T is preceded by an upturn in the ac resistivity at around 0.8 T, which evolves continuously through 0.94T. The ac resistivity is found to be roughly consistent with the dissipation measured using conventional dc transport, which does not become measurable ($> 1$ $\Omega$/sq.) until 0.97 T. Note that the resistivity being consistent between the ac and dc measurements indicates that the same underlying phenomena are being probed by the two techniques.

## IV. VORTEX CREEP AND PINNING

The films for which we show data in this manuscript have a glassy vortex state dominated by pinning and creep, and not by the melting of a disordered vortex lattice. This is most easily seen by carrying through an analysis proposed by Yazdani and coworkers [5, 6] in examining the crossover between these two behaviors. On sweeping the temperature, there is a peak in the dissipative channel ($\Re(\omega G)$), which, in a magnetic field, can be identified with a characteristic temperature scale $T_P$ for vortex motion, the origin of the dissipation. Tracking this characteristic temperature $T_P$ as a function of frequency at a fixed field, they find evidence for melting behavior at high frequencies (short length scales) in clean films, and

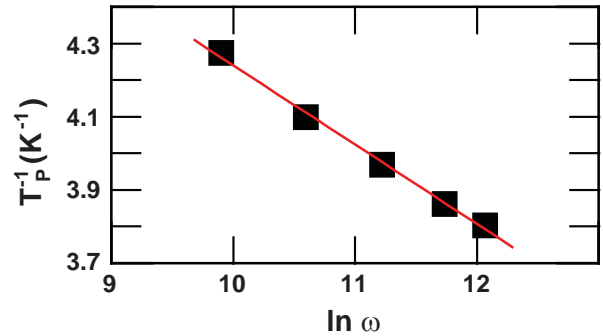

FIG. S4. Frequency dependence in finite field. The plot shows $T_P^{-1}$ plotted against the log of the measurement frequency for the MoGe film at 100 mT. The red line is a fit to the activated behavior discussed in the text.

activated behavior, which can be identified with the creep of a glassy vortex phase, in disordered films. In Figure S4, we show the results of such an analysis for the MoGe film at 100 mT. We find that this characteristic temperature follows an activated form $1/T_p = -\frac{1}{U}(ln(\omega) - ln(\omega_0))$, and hence can conclude that the vortex state of the film measured here is glassy, and dominated by creep over the length scales probed by our measurement, and not melting. This observation is consistent with the highly disordered nature of this film, where the zero-field transition is seen at a much lower temperature (0.5 K) than in thick clean films (1.05 K).

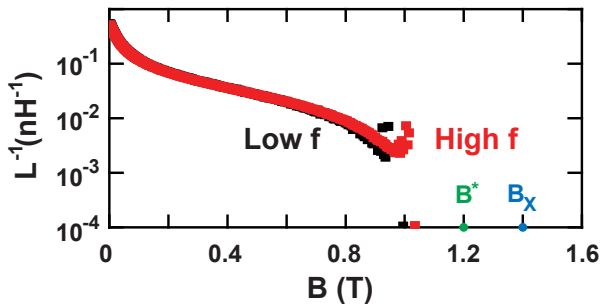

FIG. S5. Frequency dependence at low temperature. The inverse sheet inductance $L^{-1}$ of the MoGe film is shown as a function of magnetic field at a temperature of 30 mK measured at 50 kHz (black squares) and 250 kHz (red squares).The crossing field from dc resistivity isotherms ($B_X$) is marked by the blue dot, and the critical field ($B^*$) extrapolated from the 20 kHz data presented in the manuscript is marked by the green dot.

## V. FREQUENCY DEPENDENCE OF $L^{-1}$ NEAR THE FIELD-TUNED TRANSITION

The inverse sheet inductance is frequency dependent on sweeping the field at fixed temperature, as shown for our lowest temperature in Figure S5. This can be understood in terms of fluctuations near a quantum phase transition. At a low, fixed temperature, we expect superconducting correlations to appear at shorter length scales farther away from a transition, and at longer length scales closer to it. In general, data taken at lower frequency probes the system on a longer length scale than data taken at higher frequency, and thus, at a fixed temperature, the discontinuous jump in $L^{-1}$ will occur at a lower magnetic field when measured at a lower frequency. In the zero-frequency limit, this implies that, if anything, the superconducting phase terminates at fields even farther away from $B_X$ than $B^*$.

## VI. METHODS

The data reported in this manuscript were taken on two different thin films in a top- loading dilution refrigerator with a base temperature of 25 mK and a maximum field of 6 T. The amorphous $Mo_{43}Ge_{57}$ thin film (100 $\mathring{A}$) was sputtered onto a $SiO_2$-terminated Si substrate [7]. Conventional electrical transport measurements taken at zero applied magnetic field showed a normal state resistivity of $R_n = 520$ $\Omega$/sq. and a resistive transition at $T_c$ = 547 mK (Figure 1a). We also report data taken on amorphous 200 $\mathring{A}$ thin $InO_x$ films deposited on to Si substrates using an electron gun and an $In_2O_3$ [8]. The $InO_x$ film had a $T = 4$ K resistivity of 4.25 k$\Omega$/sq. The conventional electrical transport measurements were made using standard ac lockin techniques ($f = 4.87$ Hz with 1 nA of current) in a Van der Pauw geometry in separate cool downs of the same 2.5 cm square samples used for the ac impedance experiments. For the ac impedance measurements, great care was taken to ensure that the measurement was in the linear response regime throughout the phase space of temperature and magnetic field. Temperature sweeps were all taken slowly enough that no hysteresis was observed in the coil response to within the error bars used for quoting T* (typically 2-10 mK).

---


\end{document}